\documentclass[aps,twocolumn,superscriptaddress]{revtex4}%
\usepackage{graphicx}
\usepackage{amsmath}
\usepackage{amsfonts}
\usepackage{amssymb}
\usepackage{units}%
\providecommand{\U}[1]{\protect\rule{.1in}{.1in}}
\begin{document}
\title{Phase noise measurement of external cavity diode lasers and implications for
optomechanical sideband cooling of GHz mechanical modes}
\author{T.\ J.\ Kippenberg}
\email{tobias.kippenberg@epfl.ch}
\affiliation{{E}cole Polytechnique F\'{e}d\'{e}rale de Lausanne (EPFL), CH~1015,Lausanne, Switzerland}
\affiliation{Max-Planck-Institut fuer Quantenoptik, 85748 Garching, Germany}
\author{A.\ Schliesser}
\affiliation{{E}cole Polytechnique F\'{e}d\'{e}rale de Lausanne (EPFL), CH~1015,Lausanne, Switzerland}
\affiliation{Max-Planck-Institut fuer Quantenoptik, 85748 Garching, Germany}
\author{M.\ L.\ Gorodetsky}
\affiliation{Faculty of Physics, Moscow State University, Moscow 119899, Russia}

\begin{abstract}
Cavity opto-mechanical cooling via radiation pressure dynamical backaction
enables ground state cooling of mechanical oscillators, provided the laser
exhibits sufficiently low phase noise. Here, we investigate and measure the
excess phase noise of widely tunable external cavity diode lasers, which have
been used in a range of recent nano-optomechanical experiments, including ground-state cooling.
We report significant excess frequency noise, with peak values on the order of $10^{7}\,\mathrm{rad^{2}\, Hz}$ near 3.5 GHz, attributed to the diode lasers' relaxation oscillations.
The measurements reveal that even at GHz
frequencies diode lasers do not exhibit quantum limited performance.
The associated excess backaction can preclude ground-state cooling even in 
state-of-the-art nano-optomechanical systems.

\end{abstract}
\maketitle

\textit{Introduction:} 
In recent years the mutual coupling of optical and
mechanical degree of freedom has been observed in a plethora of systems and gives rise to a variety of phenomena
\cite{ Kippenberg2008, Marquardt2009, Favero2009a,Aspelmeyer2010}.
This parametric radiation pressure coupling \cite{Braginsky1977} enables sensitive measurements of the
mechanical oscillator's position, 
amplification and cooling
of mechanical motion via dynamical backaction,
optomechanical normal mode
splitting,  optomechanically induced transparency 
quantum coherent coupling of optical and mechanical degrees of freedom,
and optomechanical entanglement.
Of particular attention has been the objective to achieve ground state
cooling of a macroscopic mechanical oscillator using the technique of
optomechanical resolved sideband cooling
\cite{Schliesser2008, Marquardt2007,Wilson-Rae2007}.

Previous experiments and theoretical analysis 
\cite{
Schliesser2008, 
Diosi2008,
Rabl2009, 
Yin2009, 
Phelps2011,
Abdi2011,
Ghobadi2011}
have shown, however, that optomechanical experiments in general, and sideband cooling in particular, are sensitive to excess phase noise of the employed laser.
This necessitates the use of filtering cavities \cite{Arcizet2006a} or low-noise solid-state lasers \cite{Schliesser2008}
such as Ti:Sa and YAG\ lasers, which offer quantum-limited performance for sufficiently high Fourier frequencies (typically $>$ 10 MHz).
Diode lasers, in contrast, exhibit significant excess phase noise in this frequency range \cite{Zhang1995} and its impact has been observed in optomechanical cooling experiments of a 75 MHz radial breathing mode \cite{Schliesser2008}. 
Moreover, there exists an additional, well-known contribution \cite{Wieman1991}
to the  excess phase and amplitude noise at high Fourier frequencies ($>1\,\mathrm{GHz}$), which is fundamentally linked to damped relaxation oscillations caused by the carrier population dynamics 
\cite{Piazzolla1982, Vahala1983, Vahala1983b, Yamamoto1983}.
These relaxation oscillations cause primarily excess phase noise, whose magnitude is in close agreement with theoretical modeling  \cite{Ritter1993, Ahmed2004}.
Interestingly, optical feedback (such as provided by an external cavity)---while reducing noise at low frequency---can even lead to an enhancement of this relaxation oscillation noise \cite{Chen1984}. 

Quantitative measurements of the high frequency excess phase noise (at GHz frequencies) for modern widely tunable external cavity diode lasers, however, are scarce.
Such studies have become increasingly important as novel nano-optomechanical systems such as 1-D nanobeams \cite{Eichenfield2009} and 2-D photonic crystals \cite{Gavartin2011} operate in this GHz frequency range, and quantitative knowledge of the phase noise is therefore relevant to quantum cavity optomechanical experiments. In particular ground state cooling of a nanomechanical oscillator has been reported with an unfiltered external cavity diode laser \cite{Chan2011}. 
As such, characterization of extended cavity diode laser phase noise in the GHz domain and evaluation of its impact on quantum optomechanical experiments is highly desirable. 
Here we present such a characterization of widely tunable external cavity diode lasers
as used in recent optomechanical experiments \cite{Chan2011}.
Our results indicate that as expected, significant excess phase noise is indeed present in such lasers at GHz frequencies whose magnitude can impact optomechanical sideband cooling of nano-optomechanical systems.

\begin{figure}[b]
\centerline{ \includegraphics[width=1\linewidth]{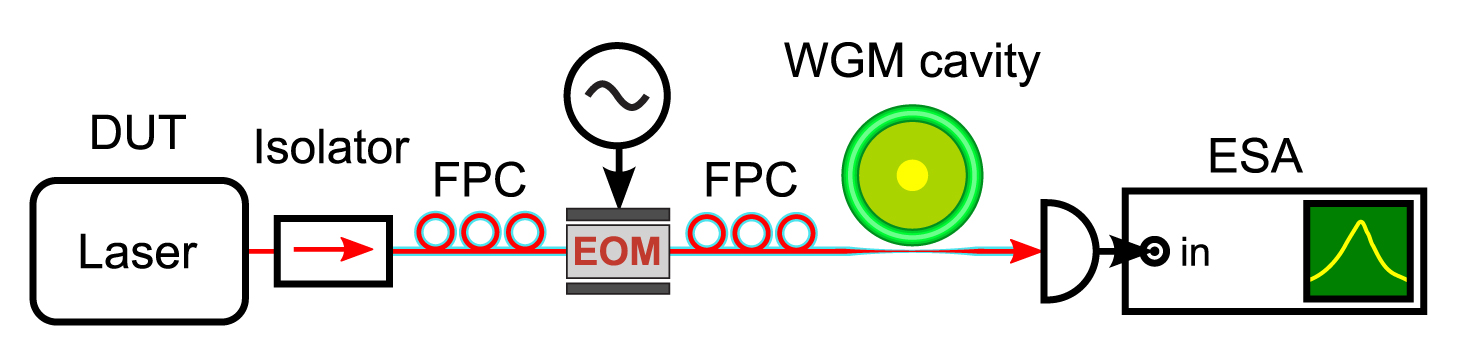} }\caption{Setup to
measure diode laser phase noise at GHz frequencies.
DUT: device under test, 
FPC: fiber polarization controller,
EOM: electro-optic modulator,
WGM: whispering gallery mode,
ESA: electronic spectrum analyzer
\label{f:setup}}%
\end{figure}

\textit{Theory: }
Radiation pressure optomechanical sideband cooling 
allows cooling of a mechanical oscillator to a minimum
occupation \cite{Wilson-Rae2007, Marquardt2007} of
$\bar n_\mathrm{f}={\kappa^{2}}/{16 \Omega_\mathrm{m}^{2}}\ll1$, where
$\Omega_\mathrm{m}$ is the mechanical frequency and $\kappa$ is
the optical energy decay rate.
This limit arises from the quantum fluctuation of the cooling laser field. 
Excess classical phase or amplitude noise causes
a fluctuating radiation pressure force noise that increases this residual
occupancy.
Of particular relevance is phase noise, whose heating effect has been
observed in experiments employing toroidal opto-mechanical resonators  \cite{Schliesser2008}.

It is instructive to first consider a coherent phase modulation at a frequency $\Omega_\mathrm{m}$ of
the lasers input field,  $s_\mathrm{in}(t)\approx 
(1+\frac {\delta\phi}{2}e^{-i\Omega_\mathrm{m} t}-\frac{\delta\phi}{2}e^{+i\Omega_\mathrm{m} t})
s_\mathrm{in}e^{-i\omega t}
$.
Pumping an opto-mechanical system residing in the resolved-sideband regime
$(\Omega_\mathrm{m}\gg\kappa)$ at the lower sideband ($\Delta=-\Omega_\mathrm{m}$)
yields an intracavity field of 
$a(t)\approx(
\frac{1}{i\Omega_\mathrm{m}}+\frac{\delta\phi}{2}\frac{1}{\kappa/2}e^{-i\Omega_\mathrm{m} t}
)
\sqrt{\eta \kappa}
s_\mathrm{in}
e^{-i\omega t}$,
where $\eta=\kappa_\mathrm{ex}/\kappa$ denotes the ratio of cavity coupling $\kappa_\mathrm{ex}$ to its feeding mode compared to total cavity losses $\kappa$.
Note that the modulation sideband is resonantly \emph{enhanced} by the cavity instead of being suppressed by a putative cavity filtering effect.
The simultaneous presence of carrier and modulation sideband leads to a radiation-pressure force 
$
F(t)
=\hbar G | a(t)|^2=
\frac{G P}{\omega}\frac{2\eta \delta\phi}{\Omega_\mathrm{m}}\sin(\Omega_\mathrm{m} t) + \bar F$,
where 
$\bar F$ is a (for the present analysis irrelevant) static force,
$P=\hbar \omega |s_\mathrm{in}|^{2}$ the launched input power
 and $G=\partial \omega_\mathrm{c}/\partial x$ is the frequency pull parameter of the optomechanical system.

These considerations carry over directly to (pure) phase fluctuations of the cooling laser field described by a (symmetrized, double-sided) spectral density $\bar S_{\phi\phi}(\Omega)$.
Alternatively, such fluctuations may be described in terms of laser frequency noise with a spectrum
$\bar S_{\omega\omega}(\Omega)=\bar S_{\dot\phi\dot\phi}(\Omega)=\bar S_{\phi\phi}(\Omega)\cdot\Omega^2$, and we use both descriptions interchangeably.
The resulting force fluctuation spectrum is given by \cite{Schliesser2008}
\begin{equation}
\bar{S}_{FF}^{\mathrm{L}}(\Omega)
\approx\frac{4\eta^{2} G^{2}P^{2}}{\omega^{2}\Omega^2} \frac{\bar{S}_{\omega\omega}(\Omega)
}{\Omega^2}%
\label{e:forcenoise}
\end{equation}
in the resolved-sideband regime.
It is straightforward to derive from this excess force noise the residual occupation of the mechanical oscillator by expressing it as an effective occupancy $\bar n_\mathrm{L}$ 
of the cold bath that the laser field is providing,
$\bar{n}_\mathrm{L}\approx
\bar{S}^\mathrm{L}_{FF}(\Omega_\mathrm{m})/2m_\mathrm{eff} \Gamma_\mathrm{m} \hbar \Omega_\mathrm{m}$, by comparing it to the Langevin force fluctuations of the thermal bath
$\bar{S}^\mathrm{th}_{FF}(\Omega)= 2m_\mathrm{eff} \Gamma_\mathrm{m} \bar n_\mathrm{th}\hbar \Omega_\mathrm{m}$
with $\bar n_\mathrm{th}=k_\mathrm{B}T /\hbar \Omega_\mathrm{m}$.
 The final occupancy of the oscillator in the presence of sideband cooling is then given by
$ n_\mathrm{f}\approx
 \left (\bar n_\mathrm{L}+ \bar n_\mathrm{th} \right)
{\Gamma_\mathrm{m}}/{\Gamma_\mathrm{cool}},
$ with 
 $\Gamma_\mathrm{cool}\approx2 \eta G^2 P/m_\mathrm{eff} \Omega_\mathrm{m}^3 \omega$
 in the resolved-sideband regime.
 This yields an excess occupancy due to frequency noise of \cite{Schliesser2008, Rabl2009}
 \begin{equation}
 \bar{n}_\mathrm{f}^\mathrm{excess}
 \approx
\frac{\bar n_\mathrm{p}}{\kappa} \bar S_{\omega\omega}(\Omega_\mathrm{m}),
\end{equation}
where $\bar n_\mathrm{p}\approx \eta \kappa P/\hbar \omega \Omega_\mathrm{m}^2$ is the intracavity photon number in the resolved-sideband regime.
For an optimized power, the lowest occupancy that can be reached is given by 
  \begin{equation}
 \bar n_\mathrm{f}^\mathrm{min}\approx\sqrt{\frac{\bar n_\mathrm{th} \Gamma_\mathrm{m}}{g_0^2} \bar S_{\omega\omega}(\Omega_\mathrm{m})},
 \end{equation}
where we have used the vacuum optomechanical coupling rate \cite{Gorodetsky2010}
$g_0=G\sqrt{\hbar/2 m_\mathrm{eff} \Omega_\mathrm{m}}$
and neglected quantum backaction
\endnote{For comparison with ref.\ \cite{Rabl2009} note that we use energy decay rates $\kappa$ instead of field decay rates
$\kappa'\equiv \kappa/2$,
and we denote with $\Gamma_\mathrm{m}$ the mechanical energy dissipation rate instead of the mechanical decoherence rate 
$\Gamma'_\mathrm{m}\equiv(k_\mathrm{B} T/\hbar \Omega_\mathrm{m}) \Gamma_\mathrm{m}$, which we refer to as $\gamma$ here.
}.

\textit{Measurement of the diode laser phase noise: }
Laser phase noise is frequently modeled by a (Gaussian) random phase $\phi(t)$ which obeys the simple noise model
$\langle \dot \phi(t) \dot \phi(s)\rangle=\gamma_\mathrm{c} \Gamma_\mathrm{L}e^{-\gamma_\mathrm{c}|t-s|}$, where 
$\Gamma_\mathrm{L}$ is the laser linewidth, and
$\gamma_\mathrm{c}^{-1}$ is a correlation time, leading to a low-pass-type
frequency noise spectrum 
\begin{equation}
  \bar S_{\omega\omega}(\Omega)=\frac{2 \Gamma_\mathrm{L} \gamma_\mathrm{c}^2}{\Omega^2+\gamma_\mathrm{c}^2}
\end{equation}
with a white noise model in the limit $\gamma_\mathrm{c}\rightarrow\infty$ \cite{Diosi2008, Rabl2009, Phelps2011, Ghobadi2011}.
In practice, the relation between the laser linewidth and the frequency noise spectrum does not follow this simple model, as there are several contributions of different physical origin to the phase noise of a diode laser:
The laser's linewidth is mostly dominated by acoustic fluctuations occurring at low Fourier frequencies, leading to a typical short-term linewidth of  $\sim 300\,\mathrm{kHz}$ for unstabilized external-cavity diode lasers.
Moreover, relaxation oscillations occur at high ($>1\,\mathrm{GHz}$) Fourier frequencies, which are not described by the above model.
Therefore, it is important to measure the frequency-dependent phase noise spectrum
$\bar S_{\phi\phi}(\Omega)$.

To this end, an optical cavity
is employed for quadrature rotation \cite{Zhang1995, Riehle2004},
converting phase to amplitude fluctuations which are measured with a photodetector (cf.\ Figure \ref{f:setup}).
In principle, a high-resolution spectrum of the optical field can also be used for phase noise measurement, in
which case the relaxation oscillations appear as sidebands around the carrier
(cf.\ e.g.\ \cite{Ritter1993, Riehle2004}).

The devices under test are three $1550\,\mathrm{nm}$
extended cavity diode lasers in Littman-Metcalf configuration of the most commonly used models
\endnote{New Focus TLB-6328 (serial numbers 008 and 280) and TLB-6330 (serial number 043).}.
Care is taken to
introduce proper optical isolation of the laser diode to avoid optical
feedback.
 The quadrature rotating cavity is a fiber coupled silica microcavity
(linewidth of $\kappa/2\pi\approx2$ $%
\operatorname{GHz}%
$) and the transmission is detected by a fast photodetector (New Focus) whose
photocurrent is fed into a spectrum analyzer (ESA).
The transduction of
frequency noise $\bar{S}_{\omega\omega}(\Omega)$ into power fluctuations at
the output of the cavity ($I$ denotes photon flux) is given by:%
\begin{align}
&  (\hbar\omega)^{2}\bar{S}_{II}^{\mathrm{fn}}(\Omega)=\label{transduction}\\
&  =\frac{4(\hbar\omega)^{2}|s_{\mathrm{in}}|^{4}\eta^{2}\Delta^{2}\kappa
^{2}\left(  (1-\eta)^{2}\kappa^{2}+\Omega^{2}\right)  \bar{S}_{\omega\omega
}(\Omega)}{\left(  \Delta^{2}+(\frac{\kappa}{2})^{2}\right)  \left(
(\Delta-\Omega)^{2}+(\frac{\kappa}{2})^{2}\right)  \left(  (\Delta+\Omega
)^{2}+(\frac{\kappa}{2})^{2}\right)  }\nonumber
\end{align}
Here, $\Delta$ is the laser detuning from the cavity resonance, and 
$\Omega$ the analysis frequency.

\begin{figure}[tb]
\centerline{ \includegraphics[width=.8\linewidth]{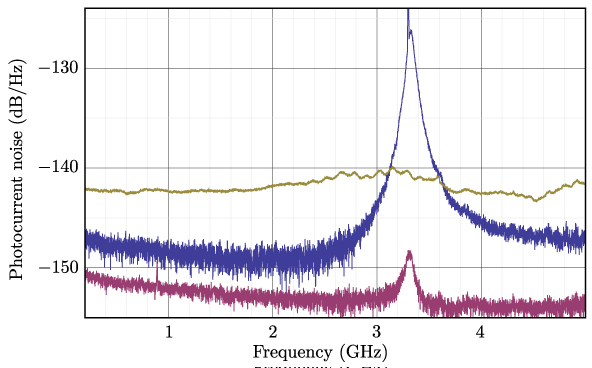} }
\caption{
Noise of a semiconductor laser with weak optical feedback from a grating
(Littman configuration). 
Shown is the power spectral density (PSD) of
photocurrent fluctuations, normalized to total photocurrent, when laser light
is directly detected (red), or tuned to the side-of-the-fringe of a
ca.\ 4\,GHz-wide optical cavity (blue). The yellow trace is the background
signal in the same normalization, which was subtracted from all traces. 
\label{f:single}
 }%
\end{figure}

\begin{figure}[b]
\centerline{ \includegraphics[width=.7\linewidth]{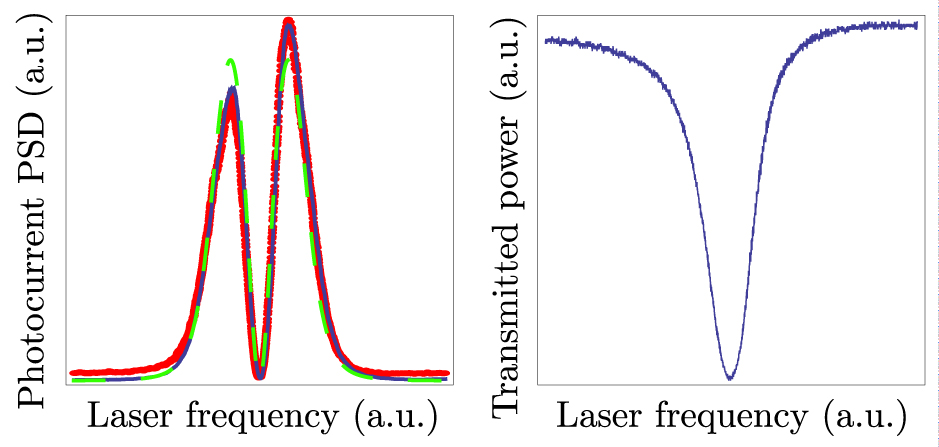}
}
\caption{%
Photocurrent PSD in a 1-MHz bandwidth at a Fourier frequency of 3.5
GHz as a function of laser detuning (left panel). 
Red dots are measured without
any additional modulation, showing only the laser's intrinsic fluctuations;
the blue line was measured with a strong external frequency modulation at 3.5
GHz.
The blue curve was rescaled by a factor of 44 and corresponds, in this
normalization, to a frequency modulation PSD of $\bar S_{\omega\omega}%
(2\pi\,3.5\,\mathrm{GHz})\approx1.6\cdot10^{7}\,\mathrm{rad^{2}\, Hz}$.
The
most striking deviation of these measurements from the model of eq.\ (\ref{transduction}) (green dashed
line) is the asymmetry of the peaks, which can be explained from the
asymmetric lineshape of the employed cavity (right panel). 
\label{f:scan}}%
\end{figure}

The noise equivalent power of the
employed photodetector is $\sim24$ p$\operatorname{W}/\sqrt
{\operatorname{Hz}}$ and therefore not sufficient to detect the quantum phase/amplitude
noise for the power levels used in this work ($< 1 \,\mathrm{mW}$),
but does allow to detect excess noise.
Indeed, as shown in Fig.\ \ref{f:single}, we
observe a peak at ca.\
$3.5\,\mathrm{GHz}$ in the detected photocurrent fluctuations when the laser is detuned from the cavity resonance.
This noise has been reported previously 
\cite{Piazzolla1982, Vahala1983, Vahala1983b, Yamamoto1983,Ritter1993, Ahmed2004} 
and is attributed to relaxation oscillations, which due to the short carrier lifetime exhibit high
frequencies well into GHz range.
We confirmed that the noise is indeed predominantly phase noise, by scanning the laser across the cavity resonance
while keeping the analysis frequency fixed (Fig.\ \ref{f:scan}).
 The pronounced
double-peak structure follows eq.\ (\ref{transduction})  and reveals that the noise is predominantly in the phase quadrature.

\begin{figure}[ptbh]
\centerline{ \includegraphics[width=1\linewidth]{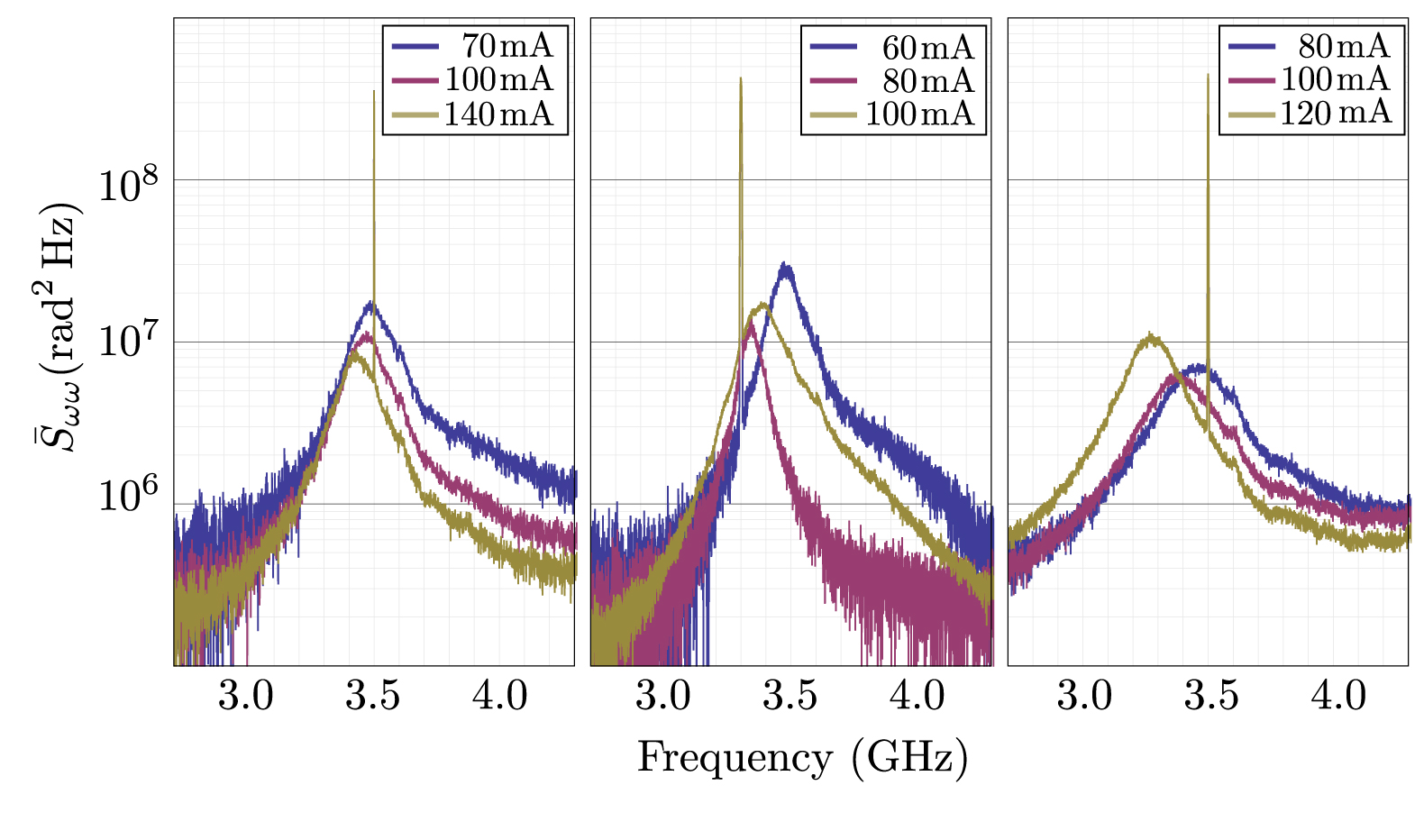}
}\caption{PSD of frequency fluctuations for three different lasers after subtraction of the background signal. 
The sharp peak is due to external phase modulation, which was used to calibrate the
spectra. \label{f:psds}}%
\end{figure}

To calibrate the measured noise spectra, we imprint onto the diode laser a
known phase modulation using an external (fiber based) phase modulator
following the method of ref.\ \cite{Gorodetsky2010}.
In brief, the $V_{\pi}$ of the phase modulator
is determined in independent measurements by scanning a second diode laser
over the phase modulated laser, in order to determine the strength of the
modulation sidebands. 
The measured $V_{\pi}$ and the manufacturers
specifications differed by typically less than 10\%.
In a second and
independent measurement (to characterize the noise level of a third laser) the
phase modulator was characterized by scanning a phase modulated laser over a
narrow cavity resonance and recording the transmission spectrum.
Calibration
via the modulation peak proceeds by using the relation $\bar S_{\omega\omega
}^\mathrm{cal}(\Omega)\equiv\Omega^{2}{\delta\phi^{2}}/{(4\cdot \mathrm{RSB})}$, where RSB is the
resolution bandwidth of the recording with the electronic spectrum analyzer, 
$\delta\phi$ the modulation index and $\Omega$ the modulation frequency.
Figure~\ref{f:psds} shows this calibration
procedure applied to the three lasers.
The level of frequency fluctuations was measured
for a total of three devices and found to vary only slightly between the lasers
(despite their differing by 10 years in manufacturing date). 
The maximum frequency noise was in the range of 
$\bar S_{\omega\omega}^{\max}\approx \mathcal{O}(10^{7})\operatorname{rad}^{2}\operatorname{Hz}$,
corresponding to phase fluctuations about $30\,\text{dB}$ above the quantum noise limit 
$\bar S_{\phi\phi}(\Omega)={\hbar \omega}/{4 P}$ of a $P=1\, \textrm{mW}$ beam.
This level of phase noise agrees well with theoretical predictions \cite{Ahmed2004}.

\textit{Ground-state cooling limitations:} 
In order to achieve ground state cooling, only a certain amount of laser phase noise can be tolerated, as the presence of the cooling light in the cavity leads to additional fluctuating forces and an excess phonon number according to eq.\ 
(\ref{e:forcenoise}).
For an optimum cooling laser power the residual thermal occupancy and the excess occupancy caused by radiation pressure fluctuations are equal, and their sum can be below unity only if %
\begin{equation}
\bar{S}_{\omega\omega}(\Omega_\mathrm{m})
<\frac{g_{0}^{2}}{\gamma}
=\frac{g_{0}^{2}}{k_\mathrm{B} T/\hbar Q_\mathrm{m}},
\end{equation}
constituting a necessary condition for ground state cooling  ($n_\mathrm{f}^\mathrm{min}<1$) \cite{Rabl2009}.
Evidently, systems that exhibit large optomechanical coupling $g_0$ and low mechanical decoherence rate
$\gamma=k_\mathrm{B} T/\hbar Q_\mathrm{m}$ (that is, low bath temperature $T$ and high mechanical quality factor $Q_\mathrm{m}$) can tolerate larger amounts of laser frequency noise.
However, even the recently reported nano-optomechanical system \cite{Chan2011} with the record-high $g_{0}
/2\pi=0.91 $ $\operatorname{MHz}$ as well as $T\approx 30\,\mathrm{K}$ and $Q_\mathrm{m}\approx50,000$ requires
$\bar{S}_{\omega\omega}(2\pi\,3.68\,\mathrm{GHz})<4\cdot10^5 \,\mathrm{rad}^2\mathrm{Hz}$, a value reached by neither of the three lasers we have tested.

We conclude that widely employed frequency-tunable external cavity diode lasers should not be expected to be quantum limited, but exhibit
significant excess phase noise up to very high Fourier frequencies.
We have observed a peak in this noise at a frequency around $3.5\,\mathrm{GHz}$.
This observation implies important limitations to optomechanical sideband cooling also for systems based on microwave-frequency mechanical oscillators if the laser noise is not suppressed, e.g.\ by external cavity filters \cite{Hald2005}.

\textit{Acknowledgements:} 
We are grateful to V.\, L.\, Velichansky and P.\ Poizat for stimulating discussions.
This work is supported by the NCCR of Quantum Engineering.

\bibliographystyle{OL}

\bibliography{/Users/aschlies/Documents/Literature/microCavities}

\end{document}